\documentstyle[12pt,epsf]{article}

  \def\pp{{\mathchoice
          {
              \kern 1pt%
              \raise 1pt
              \vbox{\hrule width5pt height0.4pt depth0pt
                    \kern -2pt
                    \hbox{\kern 2.3pt
                          \vrule width0.4pt height6pt depth0pt
                          }
                    \kern -2pt
                    \hrule width5pt height0.4pt depth0pt}%
                    \kern 1pt
           }
            {
              \kern 1pt%
              \raise 1pt
              \vbox{\hrule width4.3pt height0.4pt depth0pt
                    \kern -1.8pt
                    \hbox{\kern 1.95pt
                          \vrule width0.4pt height5.4pt depth0pt
                          }
                    \kern -1.8pt
                    \hrule width4.3pt height0.4pt depth0pt}%
                    \kern 1pt
            }
            {
              \kern 0.5pt%
              \raise 1pt
              \vbox{\hrule width4.0pt height0.3pt depth0pt
                    \kern -1.9pt  
                    \hbox{\kern 1.85pt
                          \vrule width0.3pt height5.7pt depth0pt
                          }
                    \kern -1.9pt
                    \hrule width4.0pt height0.3pt depth0pt}%
                    \kern 0.5pt
            }
            {
              \kern 0.5pt%
              \raise 1pt
              \vbox{\hrule width3.6pt height0.3pt depth0pt
                    \kern -1.5pt
                    \hbox{\kern 1.65pt
                          \vrule width0.3pt height4.5pt depth0pt
                          }
                    \kern -1.5pt
                    \hrule width3.6pt height0.3pt depth0pt}%
                    \kern 0.5pt
            }
        }}

  \def\mm{{\mathchoice
                       {
                             \kern 1pt
               \raise 1pt    \vbox{\hrule width5pt height0.4pt depth0pt
                                  \kern 2pt
                                  \hrule width5pt height0.4pt depth0pt}
                             \kern 1pt}
                       {
                            \kern 1pt
               \raise 1pt \vbox{\hrule width4.3pt height0.4pt depth0pt
                                  \kern 1.8pt
                                  \hrule width4.3pt height0.4pt depth0pt}
                             \kern 1pt}
                       {
                            \kern 0.5pt
               \raise 1pt
                            \vbox{\hrule width4.0pt height0.3pt depth0pt
                                  \kern 1.9pt
                                  \hrule width4.0pt height0.3pt depth0pt}
                            \kern 1pt}
                       {
                           \kern 0.5pt
             \raise 1pt  \vbox{\hrule width3.6pt height0.3pt depth0pt
                                  \kern 1.5pt
                                  \hrule width3.6pt height0.3pt depth0pt}
                           \kern 0.5pt}
                       }}

\catcode`@=11
\def\un#1{\relax\ifmmode\@@underline#1\else
        $\@@underline{\hbox{#1}}$\relax\fi}
\catcode`@=12

\let\du=\du                     

\def\a{\alpha}
\def\b{\beta}

\def\d{\delta}

\def\f{\phi}

\def\j{\psi}

\def\l{\lambda}

\def\q{\theta}

\def\s{\sigma}

\def\F{\Phi}

\def\ve{\varepsilon}

\def\bo{{\raise-.5ex\hbox{\large$\Box$}}}               
\def\pa{\partial}                                       
\def\TH{{\raise.2ex\hbox{$\displaystyle \bigodot$}\mskip-4.7mu \llap H \;}}
\def\face{{\raise.2ex\hbox{$\displaystyle \bigodot$}\mskip-2.2mu \llap {$\ddot
        \smile$}}}                                      

\def\Bar#1{\overline{#1}}                       
\def\leftrightarrowfill{$\mathsurround=0pt \mathord\leftarrow \mkern-6mu
        \cleaders\hbox{$\mkern-2mu \mathord- \mkern-2mu$}\hfill
        \mkern-6mu \mathord\rightarrow$}
\def\dvec#1{\vbox{\ialign{##\crcr
        \leftrightarrowfill\crcr\noalign{\kern-1pt\nointerlineskip}
        $\hfil\displaystyle{#1}\hfil$\crcr}}}           
\def\dt#1{{\buildrel {\hbox{\LARGE .}} \over {#1}}}     

\def\frac#1#2{{\textstyle{#1\over\vphantom2\smash{\raise.20ex
        \hbox{$\scriptstyle{#2}$}}}}}                   
\def\sfrac#1#2{{\vphantom1\smash{\lower.5ex\hbox{\small$#1$}}\over
        \vphantom1\smash{\raise.4ex\hbox{\small$#2$}}}} 
\def\bfrac#1#2{{\vphantom1\smash{\lower.5ex\hbox{$#1$}}\over
        \vphantom1\smash{\raise.3ex\hbox{$#2$}}}}       
\def\afrac#1#2{{\vphantom1\smash{\lower.5ex\hbox{$#1$}}\over#2}}    
\def\on#1#2{\mathop{\null#2}\limits^{#1}}               
\def\bvec#1{\on\leftarrow{#1}}                  

\def\[{\lfloor{\hskip 0.35pt}\!\!\!\lceil}
\def\]{\rfloor{\hskip 0.35pt}\!\!\!\rceil}

\def\du#1#2{_{#1}{}^{#2}}

\def\fracm#1#2{\hbox{\large{${\frac{{#1}}{{#2}}}$}}}
\def\ha{{\fracmm12}}

\def\un{\underline}
\def\fracmm#1#2{{{#1}\over{#2}}}

\def\low#1{{\raise -3pt\hbox{${\hskip 0.75pt}\!_{#1}$}}}

\def\Dot#1{\buildrel{_{_{\hskip 0.01in}\bullet}}\over{#1}}
\def\dt#1{\Dot{#1}}

\newskip\humongous \humongous=0pt plus 1000pt minus 1000pt
\def\caja{\mathsurround=0pt}
\def\eqalign#1{\,\vcenter{\openup2\jot \caja
        \ialign{\strut \hfil$\displaystyle{##}$&$
        \displaystyle{{}##}$\hfil\crcr#1\crcr}}\,}
\newif\ifdtup

\def\ibid#1#2#3{{\it ibid.}~{\bf {#1}} (19{#2}) #3}

\parskip=\medskipamount                 
\thicklines                         

\begin{document}
\thispagestyle{empty}

{\hbox to\hsize{
\vbox{\noindent April 2004 \hfill hep-th/0404119 }}}

\noindent
\vskip1.3cm
\begin{center}

{\Large\bf BPS-type Equations in the Non-anticommutative
\vglue.1in
            N=2 Supersymmetric U(1) Gauge Theory ~\footnote{Supported 
in part by the JSPS and the Volkswagen Stiftung}}
\vglue.2in

Sergei V. Ketov~\footnote{Email address: ketov@phys.metro-u.ac.jp}
and Shin Sasaki~\footnote{Email address: shin-s@phys.metro-u.ac.jp}

{\it Department of Physics, Faculty of Science\\
     Tokyo Metropolitan University\\
     1--1 Minami-osawa, Hachioji-shi\\
     Tokyo 192--0397, Japan}
\end{center}
\vglue.2in
\begin{center}
{\Large\bf Abstract}
\end{center}

\noindent
We investigate the equations of motion in the four-dimensional 
non-anticommutative N=2 supersymmetric $U(1)$ gauge field theory, in the 
search for BPS configurations. The BPS-like equations, generalizing the 
abelian (anti)self-duality conditions, are proposed. We prove full solvability 
of our BPS-like equations, as well their consistency with the equations of 
motion. Certain restrictions on the allowed scalar field values are also 
found. Surviving supersymmetry is briefly discussed too. 

\newpage

\section{Introduction}

Supersymmetric field theories in {\it Non-AntiCommutative} (NAC) superspace 
stretch limits of the conventional supersymmetric field theories formulated in 
the standard superspace with anticommutative (Grassmann) spinor coordinates.
Unlike the usual (spacetime) non-commutative field theories, where solving 
non-commutative field equations represents a formidable task \cite{dn,lech}, 
the NAC deformations can be essentially nilpotent so that they may lead to 
solvable (though non-trivial) field equations of motion. 

To the best of our knowledge, possible non-commutativity of spacetime 
(bosonic)  coordinates $x^{m}$,
$$ \[ x^{m},x^{n}\]_{\star} =i\q^{mn}~,\eqno(1.1)$$
was first proposed (in a published paper) in 1947 \cite{sny}, though Heisenberg
was known to be privately suggesting this idea in the 30's.  
Non-anticommutativity of the fermionic (Grassmann) superspace coordinates 
$\q^{\a}$ (also known as the {\it quantum superspace}),
$$ \{ \q^{\a},\q^{\b} \}_{\star} =C^{\a\b}~,\eqno(1.2)$$
was first proposed in 1981 \cite{bs} (see also ref.~\cite{ital} for more
recent developments). The non-commutativity (1.1) is well known to appear in 
superstring theory, in the presence of a constant background of the NS-NS 
antisymmetric B-field \cite{sw}. Most recent interest to the NAC supersymmetric
 field theories is due to their relevance in describing some superstring 
effective actions in certain supergravity backgrounds \cite{bgn,sei}. 

The {\it nilpotent} quantum superspace arises when merely a chiral part of the
fermionic superspace coordinates becomes NAC, whereas bosonic superspace 
coordinates still commute (in some basis) \cite{sei}. This is only 
possible when the anti-chiral fermionic coordinates $(\bar{\q})$ are not 
complex conjugates to the chiral ones, $\bar{\q}\neq(\q)^*$, which is the case
in Euclidean or Atiyah-Ward spacetimes with the signature $(4,0)$ and $(2,2)$,
 respectively. 
The Euclidean signature is relevant to instantons and superstrings \cite{sei},
whereas the Atiyah-Ward signature is relevant to the critical N=2 string 
models \cite{ov,kl} and supersymmetric self-dual gauge field theories in 2+2 
dimensions \cite{mary}. In the case of N=1 supersymmetric gauge field theories
in the nilpotent NAC superspace subject to eq.~(1.2), $(\ha,\ha)$  
supersymmetry is always broken by $C^{\a\b}\neq 0$ to $(\ha,0)$ supersymmetry,
while the change of the Lagrangian is polynomial in $C^{\a\b}$ \cite{sei}.

{\it Extended} supersymmetry is expected to bring more constraints to the NAC 
supersymmetric field theories. The NAC extension of the $N=(1,1)$ 
supersymmetric gauge field theory along the lines of ref.~\cite{sei} was 
constructed in ref.~\cite{kyoto}, with the deformed action having merely  
$(\ha,0)$ supersymmetry. There is another problem with the equations
of motion in that theory. Let's consider the NAC extension of the N=2 
supersymmetric U(1) gauge theory having the Lagrangian \cite{kyoto}
$$ \eqalign{
L ~=~ & -\fracmm{1}{4}F^{mn}F_{mn}-i\bar{\l}\bar{\s}^{m}\pa_m\l 
+\fracmm{1}{2}D^2-\fracmm{i}{2}C^{mn}F_{mn}\bar{\l}\bar{\l}\cr
&  -\pa^m\bar{A}\pa_mA -i\bar{\j}\bar{\s}^{m}\pa_m\j +\bar{F}F
 +iC^{mn}F_{mn}\bar{A}F~.}\eqno(1.3)$$
Being compared to the N=1 case represented by the first line of eq.~(1.3), the
terms in the second line are needed for N=2 extension. However, the equation 
of motion for $\bar{F}$ implies $F=0$ which, in its turn, gives rise to 
decoupling of $(A,\bar{A},\j,\bar{\j})$. The only nontrivial deformation is 
then exactly the same as in the N=1 case \cite{sei}. 

When one wants a nontrivial NAC extension of the N=2 supersymmetric gauge field
theory, one thus has to consider other deformations. The most general nilpotent
deformation of $N=(1,1)=2\times(\ha,\ha)$ supersymmetry is given by 
\cite{ilz,fs}
$$ \{ \q_i^{\a},\q_j^{\b}\}_{\star}=\d^{(\a\b)}_{(ij)}C^{(\a\b)}+
2iP\ve^{\a\b}\ve_{ij}\quad {\rm (no~sum!)}~,\eqno(1.4)$$
where $\a,\b=1,2$ are chiral spinor indices, $i,j=1,2$ are the indices of the 
internal R-symmetry group $SU(2)$, while $C^{\a\b}$ and $P$ are some constants.
Our strategy is to keep N=2 supersymmetry at the level of the NAC Lagrangian, 
and then search for its BPS-type solutions that are supposed to break some
part of supersymmetry, as is usual in field and string theories. As was noticed
in ref.~\cite{fs}, N=2 supersymmetry gives us the unique opportunity of a 
nilpotent NAC deformation preserving both N=2 supersymmetry, R-symmetry and
Lorentz invariance, when keeping only $P\neq 0$ while setting  $C^{\a\b}=0$ in
 eq.~(1.4). We believe that the NAC deformation parameter $P$ may be related to
the vacuum expectation value of some RR-type scalar in the superstring 
compactification, though we are not going to pursue the superstring connection
in this paper. Instead, we consider the equations of motion in the 
NAC-deformed N=2 supersymmetric U(1) gauge theory, and find its BPS-type 
equations, in order to demonstrate some advantages of the $P$-deformation 
versus the $C$-deformation in the case of extended supersymmetry. The 
non-abelian gauge groups in the present context will be considered elsewhere 
\cite{els}. Similar BPS-type equations in the N=1/2 supersymmetric gauge 
theories with a non-singlet NAC deformation ($C_{\a\b}\neq 0$) were derived in 
ref.~\cite{ima}.

Our paper is organized as follows. Sect.~2 is a brief review of the main 
results of ref.~\cite{fs} which is the pre-requisite to our paper. This 
section also serves as a technical introduction. In sect.~3 we establish the 
BPS-type equations in our theory and demonstrate  their consistency with the 
equations of motion. In sect.~4 we discuss a general solution to our equations
and their symmetries. Sect.~5 is our conclusion.

\vglue.2in

\section{Lagrangian}

For definiteness, we work in flat Euclidean spacetime, though we use the 
notation and conventions common to N=2 superspace with Minkowski spacetime 
signature (see ref.~\cite{book} for details about our notation). 
 
Our NAC N=2 superspace with the coordinates 
$(x^m,\q^{i}_{\a},\bar{\q}_{i}^{\dt{\a}})$ is defined by  eq.~(1.4), 
with $C^{\a\b}=0$ and $P\neq 0$, as the {\it only\/ fundamental} non-trivial  
(anti)commutator amongst the N=2 superspace coordinates. This choice preserves 
the so-called G-analyticity that is a fundamental feature of N=2 supersymmetry
\cite{fs}, while keeping the bosonic spacetime coordinates (in the G-analytic 
basis) to be commuting.

The unique star product in the NAC N=2 superspace (1.4), which preserves N=2
supersymetry, R-symmetry and `Lorentz' invariance, is given by \cite{fs}
$$ A\star B = A\exp\left( iP\ve^{\a\b}\ve^{ij}\bvec{D}{}_{i\a}\vec{D}{}_{j\b}
\right)B~~,\eqno(2.1)$$
where $D_{i\a}$ are the standard N=2 chiral supercovariant derivatives. 
The star product (2.1) allows us to introduce N=2 {\it anti-chiral} superfields
(defined by $D_{i\a}\bar{\F}=0$). In the N=2 superspace {\it anti-chiral} 
basis, the N=2 chiral supercovaraint derivatives are simply given 
by $D_{i\a} =-\pa/\pa\q^{i\a}$, while the standard abelian N=2 anti-chiral 
superfield strength in components (to be defined this way after expanding in
powers of $\bar{\q}$) reads 
$$\eqalign{
 \Bar{W}(x_R,\bar{\q}) = ~&~ \bar{\f} + \bar{\q}_{i\dt{\a}}\bar{\l}^{i\dt{\a}}
+ \fracmm{1}{8\sqrt{2}}(\bar{\q}_i\tilde{\s}^{mn}\bar{\q}^i)F^-_{mn} \cr
 ~&~ +\fracmm{1}{2\sqrt{2}}\bar{\q}_{ij}D^{ij}
 + (\bar{\q}^3)^i_{\dt{\a}}(\tilde{\s}^m)^{\dt{\a}\a}\pa_m\l_{i\a}
- (\bar{\q})^4\bo\f~~.\cr}\eqno(2.2)$$
The scalars $\bar{\f}$ and $\f$, as well as the chiral spinors 
$\bar{\l}^{i}_{\dt{\a}}$ and $\l_{i\a}$, are the {\it independent} fields 
in Euclidean or Atiyah-Ward spacetimes, i.e. they are not related by complex 
conjugation, contrary to Minkowski spacetime. The self-dual and
anti-selfdual parts of the Maxwell field strength, $F^+_{mn}$ and  $F^-_{mn}$
respectively, are defined by $F^{\pm}=\ha(F\pm {}^*F)$, where ${}^*F$ is the 
dual field strength, ${}^*F^{mn}=\ha\ve^{mnpq}F_{pq}$. The spacetime 
derivatives in eq.~(2.2) at the $(\bar{\q}^3)$ and $(\bar{\q})^4$ terms are 
needed to  provide canonical dimensions to the component fields $\l$ and $\f$,
 while they are also consistent with the Bianchi identities on $F_{mn}$ and 
N=2 supersymmetry. The $SU(2)$ triplet $D_{ij}$ are the auxiliary fields. 

The free action of the N=2 supersymmetric $U(1)$ gauge theory in the standard 
(undeformed) N=2 superspace is given by \cite{wess}:
$$  S_{\rm free} = \fracm{1}{2}\int d^4x_R d^4\bar{\q}\;\Bar{W}\,{}^2~~.
\eqno(2.3)$$
The most general deformation of the action (2.3), compatible with N=2 
supersymmetry, R-symmetry and `Lorentz invariance', and having no higher
derivatives, is parameterized by an arbitrary real function  $f(\Bar{W})$,
$$  S_{f} = \fracm{1}{2} \int d^4x_R d^4\bar{\q}\;f(\Bar{W})\equiv
\int d^4x L_f~~.\eqno(2.4)$$

It is straightforward to calculate the component Lagrangian associated with
the action (2.4). The result in our notation is given by 
$$\eqalign{
 L_f ~=~& -\fracm{1}{2}f'(\bar{\f})\bo\f -\fracm{1}{4}f''(\bar{\f})(F^-_{mn})^2
-\fracm{1}{2}f''(\bar{\f})\bar{\l}_i\tilde{\s}^m i\pa_m \l^i +
\fracm{1}{8}f''(\bar{\f})D^{ij}D_{ij} \cr
~ & - \fracmm{1}{4\sqrt{2}}f'''(\bar{\f})\bar{\l}_{i\dt{\a}}\bar{\l}_{j\dt{\b}}
\left[ \fracm{1}{4}\ve^{ij}(\tilde{\s}^{mn})^{\dt{\a}\dt{\b}}F^-_{mn}
+\ve^{\dt{\a}\dt{\b}}D^{ij}\right] 
+ \fracm{1}{2}f''''(\bar{\f})(\bar{\l})^4~,\cr}\eqno(2.5)$$
where the primes denote differentiations with respect to $\bar{\f}$.

Since the NAC $P$-deformation does not break N=2 supersymmetry, R-symmetry and
 `Lorentz invariance' either, the resulting non-linear action in 
the standard N=2 superspace must belong to the family (2.4). An actual 
calculation of the effective function $f$ requires the explicit use of the 
N=2 gauge superfield pre-potentials in the non-abelian setup, which is only 
possible in N=2 harmonic superspace. This calculation was done in 
ref.~\cite{fs} with the following result:
$$f(\Bar{W}) =\fracmm{\Bar{W}\,{}^2}{(1+P\Bar{W})^2}~,\quad {\rm 
or,~effectively,}
 \quad \Bar{W}_{\rm nac} =\fracmm{\Bar{W}}{(1+P\Bar{W})}~~.\eqno(2.6)$$
This is to be compared to the exact solution to the Seiberg-Witten map 
 \cite{sw} in the abelian (but spacetime non-commutative) case \cite{liu}:
$$ (F_{\rm nc})_{mn}= \fracmm{F_{mn}}{1+F\cdot \q}~,\quad {\rm where}\quad
F\cdot \q =\q^{mn}F_{mn}~~.\eqno(2.7)$$
Both effective functions in eqs.~(2.6) and (2.7) are apparently the same, 
though  eq.~(2.6) is Lorentz-invariant and N=2 supersymmetric whereas 
eq.~(2.7) is not. 

\section{BPS equations}

It is straightforward to calculate the Euler-Lagrange equations of motion from
the Lagrangian (2.5). The variations with respect to $\f$ and $\l^{i}$ 
give rise to the equations, respectively,
$$ \bo (f'(\bar{\f}))=0 \quad {\rm and}\quad
 i\s^m\pa_m\left(f''(\bar{\f})\bar{\l}_i\right)=0~~.\eqno(3.1)$$
The variation with respect to the abelian gauge field $A^n$ yields
$$ \pa^m\left[f''(\bar{\f})F^-_{mn}+\fracmm{1}{8\sqrt{2}}
f'''(\bar{\f})(\bar{\l}^i\tilde{\s}_{mn}\bar{\l}_i)\right]=0~.\eqno(3.2)$$
The variation with respect to $D_{ij}$ gives rise to the algebraic constraint
$$ f''(\bar{\f})D^{ij} -\fracmm{1}{\sqrt{2}}
f'''(\bar{\f})(\bar{\l}^i\bar{\l}^j)=0~~.\eqno(3.3)$$
The variation with respect to $\bar{\l}_{i\dt{\a}}$ yields the fermionic 
equation of motion 
$$ f''(\bar{\f})(i\tilde{\s}^m\pa_m\l^{i})_{\dt{\a}}
+\fracmm{1}{\sqrt{2}} f'''(\bar{\f})\bar{\l}_{\dt{\a}j}D^{ij}
 - f''''(\bar{\f})(\bar{\l}^3)^{i}_{\dt{\a}}=0~,\eqno(3.4)$$
while the variation with respect to $\bar{\f}$ leads to the bosonic equation 
of motion
$$ \eqalign{
\bo \f ~=~ & -\fracmm{f'''(\bar{\f})}{2f''(\bar{\f})}\left[
 (F^-_{mn})^2+2i(\bar{\l}_i\s^m\pa_m\l^i)-\fracmm{1}{2}D^{ij}D_{ij}\right.\cr
~ & \left. +\fracmm{f''''(\bar{\f})}{\sqrt{2}f'''(\bar{\f})}
\left\{\fracm{1}{4}(\bar{\l}^i\tilde{\s}^{mn}\bar{\l}_i)F^-_{mn}
+ (\bar{\l}_i\bar{\l}_j)D^{ij}\right\} -
\fracmm{2f'''''(\bar{\f})}{f'''(\bar{\f})}(\bar{\l})^4 \right]~.\cr}
\eqno(3.5)$$

Equations (3.1) amount to the {\it free\/} Klein-Gordon and Dirac 
equations of motion, 
$$ \bo \bar{\f}_{\rm new}=0 \quad {\rm and} \quad  
i\s^m\pa_m\bar{\l}^i_{\rm new}=0~,\eqno(3.6)$$
after the field redefinitions 
$$ f'(\bar{\f})=\bar{\f}_{\rm new} \quad {\rm and}\quad
f''(\bar{\f})\bar{\l}^{\dt{\a}i}=\bar{\l}^{\dt{\a}i}_{\rm new}~~.\eqno(3.7)$$
  
Equations (3.3) clearly determine the auxiliary fields as follows: 
$$D^{ij}=-\fracmm{f'''(\bar{\f})}{\sqrt{2}f''(\bar{\f})}
(\bar{\l}^i\bar{\l}^j)~~.\eqno(3.8)$$
The equations of motion (3.2), (3.4) and (3.5) are non-trivial, whose 
interaction terms are entirely fixed by the NAC deformation $P\neq 0$. It is
worth mentioning that the new NAC-generated contributions (i.e. those 
vanishing at $P=0$) correspond to the terms above that contain the third or 
higher derivatives of $f$.

The Euler-Lagrange equations of motion above have the second-order
derivatives for bosons and the first-order derivatives for fermions. The BPS 
equations should be of the first order in the bosonic fields, they are supposed
to be some deformed version of the abelian selfduality or anti-selfduality
equations, they should also be derivable by minimizing the Euclidean action, 
while they are to be consistent with the Euler-Lagrange equations of motion 
too.  Since the NAC deformation is essentially chiral, the selfdual and 
anti-selfdual cases differ, so that they are to be considered separately.

\underline{NAC {\it selfduality} equations}. The Lagrangian (2.5) does not 
depend upon $F^+_{mn}$ at all, so that the undeformed self-duality condition
is the only possibility,
$$ F^+_{mn}=0~.\eqno(3.9)$$
A consistency between eqs.~(3.2) and (3.9) is only possible when 
$\bar{\f}=const.$ because the required condition 
$\pa_m(\bar{\l}_i\tilde{\s}^{mn}\bar{\l}_i)=0$ then
appears to be the consequence of the free Dirac equation of motion in 
eq.~(3.1). The spinor field $\bar{\l}_{i\dt{\a}}$ represents the undeformed 
anti-chiral fermionic zero modes in this case.

\underline{NAC {\it anti-selfduality} equations}. Arranging the perfect square
involving $F^{-}_{mn}$ in the Lagrangian (2.5) and minimizing the action give 
rise to the $P$-deformed anti-selfduality equation 
$$ f''(\bar{\f})F^-_{mn} +\fracmm{1}{8\sqrt{2}}
f'''(\bar{\f})(\bar{\l}^i\tilde{\s}_{mn}\bar{\l}_i)=0~.\eqno(3.10)$$
This is obviously consistent with the equation of motion (3.2). Substituting 
eqs.~(3.3) and (3.10) back into the Lagrangian (2.5) eliminates the terms 
quadratic in $\bar{\l}$, so that the remaining fermionic interaction terms
become quartic in $\bar{\l}$, 
$$ L_{\rm fermi-int.}= F(\bar{\f})(\bar{\l})^4~,\quad {\rm where}\quad
F(\bar{\f})=-\fracmm{3f'''(\bar{\f})^2}{2f''(\bar{\f})} 
+\fracmm{1}{2}f''''(\bar{\f})~.\eqno(3.11)$$
Similarly, the rest of equations (3.4) and (3.5) is now given by
$$ i\tilde{\s}^m\pa_m\l^i -\fracmm{2F(\bar{\f})}{f''(\bar{\f})}(\bar{\l}^3)^i=0
\eqno(3.12)$$
and
$$ \bo \f =J(\bar{\f})(\bar{\l})^4~,\quad {\rm with}\quad
J(\bar{\f})=\fracmm{4f'''(\bar{\f})}{f''(\bar{\f})^2}F(\bar{\f})-
 \fracmm{2}{f'''(\bar{\f})}\fracmm{\pa F(\bar{\f})}{\pa\bar{\f}}~.
\eqno(3.13)$$

After  taking into account eq.~(2.6) we find
$$ F(\bar{\f})= \fracmm{12P^2(5P^2\bar{\f}^2-10P\bar{\f}+6)}{(1+
P\bar{\f})^6(-1+2P\bar{\f})}\eqno(3.14)$$
and
$$ J(\bar{\f})= \fracmm{24P^3(5P^3\bar{\f}^3-15P^2\bar{\f}^2+
24P\bar{\f}-19)}{(1+P\bar{\f})^3(-1+P\bar{\f})^3}~.\eqno(3.15)$$ 

The function (3.14) becomes singular at $P\bar{\f}=-1$ and  
$P\bar{\f}=+\fracm{1}{2}$, where our effective description of NAC breaks 
down.~\footnote{We assume, for definiteness, that $P\geq 0$.} Hence, we get
limits of the allowed real (physical) values of the field $\bar{\f}$,
$$   -1 < P\bar{\f} < \fracmm{1}{2}~~.\eqno(3.16)$$

The quadratic polynomial $5P^2\bar{\f}^2-10P\bar{\f}+6$ in the numerator of the
 function $F$ in eq.~(3.14) is always positive for all real values of 
$\bar{\f}$, so that the fermionic equation (3.12) cannot become a free Dirac 
equation unless $P=0$ or $P\to\infty$. The graphs of the functions 
$F(\bar{\f})$ and $J(\bar{\f})$ are given in Fig.~1.

In the anti-commutative limit, $P\to +0$, the obstructions (3.16) disappear, 
as they should.

\begin{figure}[ht]
\begin{center}
\vglue.1in
\makebox{
\epsfxsize=4in
\epsfbox{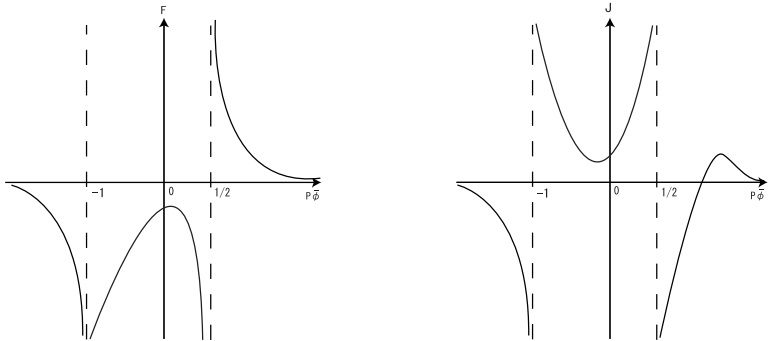}
}
\caption{\small Graphs of the functions $F$ in eq.~(3.14) and the function $J$
in eq.~(3.15).}
\end{center}
\end{figure}

\section{Solutions and Symmetries}

The remarkable fact about our BPS-like equations (sect.~3) is their 
solvability. A general solution to the field equations can be written down 
explicitly, despite of their apparently non-linear form. This seems to be the 
very special feature of our abelian NAC theory that does not seem to have an
analogue in Minkowski spacetime.

Let's begin with equations (3.1). They can be brought to the form (3.6) after 
the algebaric field redefinition (3.7). The most general solutions to the {\it 
free} Klein-Gordon and Dirac equations (3.6) are well known, either in 
Euclidean or Atiyah-Ward spacetime. Inverting the algebraic  relations (3.7) 
we can get both functions $\bar{\f}$ and $\bar{\l}$ in terms of the general
solution to eq.~(3.6). Further, eq.~(3.10) actually delivers us a solution to 
$F^-_{mn}$ in terms of the already known functions  $\bar{\f}$ and $\bar{\l}$.
Now the remaining equations (3.12) and (3.13) take the form of the Dirac and 
Klein-Gordon equations on $\l$ and $\f$, respectively, with the {\it known} 
sources. Hence, they can also be easily solved explicitly, by using the 
standard Green functions of the {\it free} Klein-Gordon and Dirac operators. 

The N=2 supersymmetry transformations of all the field components in the N=2
superfield (2.5) follow by direct calculation from N=2 superspace (see e.g., 
ref.~\cite{book}). We find 
$$ \eqalign{
\d \bar{\f} ~&~ = -\bar{\ve}_{\dt{\a}j}\bar{\l}^{\dt{\a}j}~~,\cr
\d \bar{\l}_{\dt{\a}i} ~&~ = 
\fracmm{1}{4\sqrt{2}}(\tilde{\s}^{mn})_{\dt{\a}\dt{\b}}
\bar{\ve}^{\dt{\b}}_iF^-_{mn} +\fracmm{1}{\sqrt{2}}\bar{\ve}^j_{\dt{\a}}D_{ij}
-2i\ve^{\a}_i(\s^m)_{\a\dt{\a}}\pa_m\bar{\f}~,\cr }
\eqno(4.1)$$
and similarly (by formal `complex conjugation') for $\f$ and $\l$, as well as 
$$ \d F^-_{mn} =\fracmm{-i}{\sqrt{2}}\left\{ (\ve_j\s_{[m}\pa_{n]}\bar{\l}^j)-
 (\bar{\ve}_j\tilde{\s}_{[m}\pa_{n]}\l^j) -\fracm{1}{2}\ve_{mnpq}\left[
(\ve_j\s_p\pa_q\bar{\l}^j)-(\bar{\ve}_j\tilde{\s}_p\pa_q\l^j)\right]\right\}~,
\eqno(4.2)$$
where we have introduced the infinitesimal anticommuting (Grassmann) spinor 
parameters $(\ve_{i\a},\bar{\ve}^{i\dt{\a}})$ of rigid N=2 supersymmetry. 

It is easy to see that the self-duality condition (3.9) implies $\l_{i\a}=0$
by supersymmetry, as well as $\bar{\f}={\rm const.}$, as it should have been 
expected from supersymmetry. Hence, some supersymmetry may be preserved 
in this case. 

As regards the anti-selfdual case, requiring a half of the N=2 supersymmetry 
transformations in eq.~(3.10) to vanish leads to one of following fermionic 
equations:
$$ \left(\tilde{\s}_{[m}\pa_{n]}\l^i\right)^- =
\fracmm{if'''(\bar{\f})}{8f''(\bar{\f})}\bar{\l}^{ij}
\left(\tilde{\s}_{mn}\bar{\l}_j\right)~~,\eqno(4.3a)$$
or
$$ 
\left(\s_{[m}\pa_{n]}\bar{\l}_i\right)^-=
-\fracmm{2f'''(\bar{\f})}{f''(\bar{\f})}
\left(\s_{[m}\pa_{n]}\bar{\f}\right)^-\bar{\l}_i~,\eqno(4.3b)$$
where we have used the fact that the rigid N=2 supersymmetry parameters 
$\ve_{i\a}$ and $\bar{\ve}^{i\dt{\a}}$ are independent, as well as the
notation
$$ X_{mn}^-=X_{mn}-\fracm{1}{2}\ve_{mnpq} X_{pq}\eqno(4.4)$$
to denote the anti-selfdual part of an antisymmetric tensor $X_{mn}$. The 
fermionic BPS-like conditions (4.3) are complementary to the bosonic  BPS-like 
equation (3.10), as long as some part of supersymmetry is preserved. 

The simplest solution to eq.~(4.3b) is given by $\bar{\l}^j_{\dt{\a}}=0$. This
is consistent with the equations of motion, it implies $\bar{\f}={\rm const.}$ 
and $F^-_{mn}=0$, while it preserves half of supersymmetry. Unfortunately, 
this solution does not lead to a non-trivial deformation of the abelian 
anti-selfduality equation \cite{mary}. It remains to be seen whether there are
 other non-trivial solutions to eq.~(4.3) that would be consistent with the 
equations of motion and, hence, preserve some part of supersymmetry by 
construction.

\newpage

\section{Conclusion}

Our considerations in this paper were entirely classical. It would be 
interesting to investigate the role of quantum corrections both in quantum
field theory and in string theory (e.g. by using the geometrical engineering).
In particular, the standard Seiberg-Witten solutions \cite{sws} in the Coulomb 
branch of quantum N=2 super-Yang-Mills theories formally belong to the class 
(2.4) considered in this paper, so that it is conceivable to conjecture that 
quantum corrections to the NAC solution (2.6) may be linked to the 
Seiberg-Witten theory.

Upon dimensional reduction to {\it two} spacetime dimensions, the field theory
 (2.5) with {\it any} function $f$ gives rise to the two-dimensional N=4 
supersymmetric non-linear sigma-model with a non-trivial torsion (or a 
generalized Wess-Zumino term), whose geometry and renormalization were 
investigated in ref.~\cite{it}. In particular, as was shown in ref.~\cite{it},
all those two-dimensional supersymmetruic non-linear sigma-models are always
ultra-violet {\it finite}, as the quantum field theories, in all loops.

\section*{Acknowledgements}

One of the authors (S.V.K.) acknowledges discussions with S. J. Gates Jr., 
Olaf Lechtenfeld, Andrei Losev, Andrei Mironov and Werner R\"uhl, he thanks 
Silvia Penati for correspondence, and he is also grateful to the Department of 
Physics of the University of Maryland in College Park, USA, the Insititute for
Theoretical Physics of Hannover University and the Department of Physics of 
Kaiserslautern University in Germany, for kind hospitality extended to him 
during some parts of this work.

\end{document}